\documentstyle[psfig,twocolumn,aps]{revtex}
\begin{document}

\draft
\title{Electron Spin Flip Relaxation by One Magnon Processes:\\
Application to the Gadolinium surface band}

\author{ Philip B. Allen}
\address{Department of Physics and Astronomy, State University of New York,
Stony Brook, New York 11794-3800}

\date{\today}

\maketitle

\begin{abstract}
The ``s-f model,'' also known as the ferromagnetic Kondo lattice,
contains a description of band electrons coupled to localized spins
which is an appropriate description of the magnetic part of the
low-energy physics of Gd metal.
Here the model is used to estimate the lifetime broadening of the
minority spin component of the surface electron band in ferromagnetic
gadolinium metal at temperatures below the Curie temperature.
The low temperature result $1/\tau\approx$0.1 eV agrees nicely with 
a measurement by Fedorov {\sl et al.}
\end{abstract}


\section{introduction}

Fedorov {\sl et al.} \cite{Fedorov} have recently measured the line-shape
$A(\vec{k},\omega)$ of photoemitted electrons in the
ferromagnetic metal Gd.
For photoelectrons associated with a photohole in the
$(001)$ surface band, they find somewhat different life-time
broadening depending on whether the emitted electron has up
(majority) or down (minority) spin.  They interpret the source
of lifetime broadening to be electron-phonon scattering for the majority
spin component of the photohole, and electron-magnon scattering for the 
minority spin component of the photohole.  Although the arguments 
given by Fedorov {\sl et al.}
seem perfectly sensible, nevertheless, this interesting diversity
suggests a need for theoretical inquiry.  The magnitude
of electron-phonon scattering in Gd has previously \cite{Skriver}
been estimated, with results roughly agreeing with the assigned majority spin
equilibration rate.  Electron-magnon scattering has not previously
been estimated.

Here I argue that the ``s-f'' or ``ferromagnetic
Kondo lattice'' model allows reasonable estimates without
free parameters.  I make an extreme model for the nature of
surface electron and surface magnon states:
surface electron states have amplitude 1 on the top
layer and zero elsewhere, while magnon states at the surface \cite{Mills}
are simply the bulk Bloch states, ignoring surface boundary conditions.
Using this model, and the measured mass $m^{\ast}\approx 1.2$
of the surface hole band, the zero temperature equilibration
rate of minority spin holes more than 25meV from the top of the hole band
is predicted to be $1/\tau=$0.10 eV, agreeing with the experiment.

\section{formula for relaxation rate}

The generic Hamiltonian for the ``s-f'' or ``ferromagnetic
Kondo lattice'' model couples electron bands $\vec{k}n$
with energy $\epsilon(\vec{k}n)$ (independent of spin, so far)
to localized spins $\vec{S}_i$ located on atoms at lattice
sites $\vec{R}_i$:
\begin{eqnarray}
{\cal H}&=&
\sum_{\vec{k} n \sigma} \epsilon(\vec{k}n)
c^{\dagger}(\vec{k} n \sigma)c(\vec{k} n \sigma) \nonumber \\
&-&J\sum_{i m \alpha\beta} \vec{S}_i \cdot
c^{\dagger}(i m \alpha)\vec{\sigma}_{\alpha\beta}c(i m \beta).
\label{eq:ham}
\end{eqnarray}
The electron bands derive from the outer atomic orbitals
$|im\alpha>$ with wavefunctions
$\psi_m(\vec{r}-\vec{R}_i)\chi_{\alpha}$ where $\chi_{\alpha}$ is
the spin part. This model should describe the
low-energy spin-related physics of a metal like Gd, with
three conduction electrons per atom in orbitals derived from
atomic $s$ and $d$ states of Gd.  These three orbitals have a
magnetic interaction with the well-localized $S=7/2$ half-filled
$4f$ shell.  The exchange parameter $J$ comes from the atomic
Hund's rule trying to keep ``electron'' spins parallel to
``core'' spins.
The same Hamiltonian, in the $J\rightarrow\infty$ limit, is 
known as the ``double exchange Hamiltonian''  \cite{Zener} and is very popular
right now \cite{Millis} for discussions of hole-doped LaMnO$_3$.

The zero temperature phase diagram of the s-f model (for a single $s$ band) as
a function of filling and $J/$band-width ratio has been computed
approximately \cite{Chattopadhyay}.  Ferromagnetic order occurs
over a wide range of parameters, with a Curie temperature
($T_c$=292K for Gd) proportional to $J$.  Electron bands acquire a
spin splitting proportional to $J$. Lindgard {\sl et al.}
\cite{Lindgard} used the model to calculate (in random-phase
approximation) the spin-wave dispersion, which was measured for Gd
by Koehler {\sl et al.} \cite{Koehler}, and which has been studied
using spin-density functional theory by Perlov {\sl et al.} \cite{Perlov}.

In lowest-order spin-wave approximation, we replace the
Fourier-transformed spin operator
$\vec{S}_Q$ in Eq. (\ref{eq:ham}) by spin-wave creation
and destruction operators $a_Q^{\dagger}$ and $a_Q$ using
$S_{Qz}=S\delta_{Q,0}-a_Q^{\dagger}a_Q$,
$(S_{Qx}+iS_{Qy})/2=S_Q^+=\sqrt{2S}a_Q$,
and $(S_{Qx}-iS_{Qy})/2=S_Q^-=\sqrt{2S}a_Q^{\dagger}$.
The $Q=0$ term in lowest order gives spin splitting $2JS$,
lowering the energy of bands with spin parallel to the
localized spin $\vec{S}_i \sim S\hat{z}$ and raising the other
bands equally.  The $S^{\pm}$ terms give rise to spin-flip scattering
events.

\begin{figure}[t]
\centerline{\psfig{figure=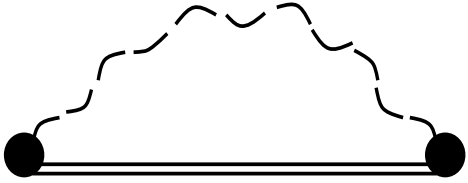,height=1.0in,width=2.4in,angle=0}}
\caption{Feynman graph for electron self-energy from electron-magnon
coupling.  The double solid line is the renormalized electron Green's
function; the wavy line is the renormalized magnon Green's function;
circles are the effective electron-magnon interaction matrix element.}
\label{fig:feynman}
\end{figure}

In this note I estimate the rate $1/\tau$ at which a single
out-of-equilibrium hole in
a surface band relaxes back toward equilibrium by spin-flip
processes.  The rate can be found from $\hbar/\tau=-2{\rm Im}\Sigma$,
where the leading Feynman diagram for the self-energy $\Sigma$ is
shown in Fig. \ref{fig:feynman}.  Only the one-magnon process is
considered.  This approximation can be questioned on the grounds
of fallibility of the ``Migdal approximation'' for electron-magnon
processes \cite{Hertz}. However, it is a proper first estimate,
the only one which can be reliably computed, and seems to me
unlikely to make a large error when $T\ll T_c$.
Two magnon processes have been considered by Lutovinov and Reizer 
\cite{Reizer}.

\par
\par

The actual states $(\vec{k},n)$ of interest are surface states of
spin {\bf primarily} but not exclusively up. The probability
$p(\vec{k},n,\uparrow)$ (or $p(\uparrow)$ for short) that the spin
state is up, is less than one.  The amplitude of the up-spin
component of the wavefunction is $\sqrt{p(\uparrow)}$ and is close
to 1.  The corresponding probability
$1-p(\uparrow)=p(\vec{k},n,\downarrow)$ (or $p(\downarrow)$ for
short) that the spin state is down, is non-zero for two reasons.
First, the spin-orbit interaction is not small and mixes spin
states. Second, $f$-spins on Gd atoms may deviate from perfect
alignment by quantum and thermal fluctuations, and the conduction
states are locally locked by Hund's rule in the same spin
orientation as the $f$-spins.  The second process is a
renormalization of electron bands by magnon processes.
A combination of the two effects is seen experimentally
\cite{Fedorov} as a small minority spin component in the
photoemitted electron.  This component is sometimes referred
to as a ``shadow band'' and has received recent theoretical
treatments \cite{MacDonald}.

An elementary derivation of the ``Migdal'' result follows from the
standard ``Golden rule'' rate equations of Boltzmann theory.
Suppose the down-spin component of state $\vec{k}n$ has population
$p(\downarrow)F(\vec{k},n,\downarrow)$. If this deviates from the
equilibrium Fermi-Dirac population
$p(\downarrow)f(\vec{k},n,\downarrow)$, then it will evolve back
toward equilibrium according to
\begin{eqnarray}
\frac{dp(\downarrow)F(\vec{k},n,\downarrow)}{dt} &=&
-\frac{2\pi}{\hbar}\sum_{\vec{Q}n^{\prime}b}
|M_b(\vec{k}n\downarrow,\vec{k}+\vec{Q}n^{\prime}\uparrow)|^2
\nonumber \\ &\times&\left\{L({\rm emission})+L({\rm
absorption})\right\}. \label{eq:grule}
\end{eqnarray}
\begin{eqnarray}
\lefteqn{L({\rm emission})=\delta(\epsilon-\epsilon^{\prime}
-\omega^{\prime})} \nonumber \\
\hspace{.5in} &\times& \left[
(N^{\prime}+1)F(1-F^{\prime})-N^{\prime}F^{\prime}(1-F) \right]
\label{eq:emission}
\end{eqnarray}
\begin{eqnarray}
\lefteqn{L({\rm absorption})=\delta(\epsilon-\epsilon^{\prime}
+\omega)} \nonumber \\
\hspace{.5in} &\times& \left[NF(1-F^{\prime})-(N+1)F^{\prime}(1-F) \right]
\label{eq:absorption}
\end{eqnarray}
Here $M_b()=J[(2S/2N)p(\downarrow)p^{\prime}(\uparrow)]^{1/2}$ is
the matrix element for the process $(\vec{k},n,\downarrow)$
scattering to $(\vec{k}+\vec{Q}n^{\prime}\uparrow)$ by emission of
the magnon $(-\vec{Q}b)$ or absorption of the magnon $(\vec{Q}b)$.
The factor $1/\sqrt{2N}$, where $N$ is the number of unit cells
and $2N$ the number of atoms in the sample, comes from
normalization of the spin-wave eigenvector.  There are two
branches of spin waves, with amplitude $\pm 1/\sqrt{2}$ on each
atom. This is discussed in the Appendix.  A shorthand is used that
$\epsilon$ and $p(\downarrow)F$ stand for the energy and occupancy
of the quasiparticle state $(\vec{k}n\downarrow)$,
$\epsilon^{\prime}$ and $p(\uparrow)F^{\prime}$ stand for the
energy and occupancy of the quasiparticle state
$(\vec{k}+\vec{Q}n^{\prime}\uparrow)$, $\omega$ and $N$ stand for
the energy and occupancy of the magnon state $(\vec{Q}b)$, and
$\omega^{\prime}$ and $N^{\prime}$ stand for the energy and
occupancy of the magnon state $(-\vec{Q}b)$. At temperature $T>0$,
depletion of an excess population toward equilibrium occurs both
by emission and absorption of thermal magnons.  Each process
(emission or absorption) has a time-reversed process which
enhances the population, the ``scattering in'' terms with opposite
sign.  In thermal equilibrium, scattering out and in occur at
equal rates.  This ``principle of detailed balance'' guarantees
that the two parts of $L$(emission) cancel each other when the
distributions $N$ and $F$ become the equilibrium distributions $n$
and $f$, and similarly for $L$(absorption).

Now make the assumption that all quasiparticles are in equilibrium
except for a particular state $(\vec{k}n\downarrow)$ of interest,
whose population ($F$) deviates from equilibrium ($f$) by $\delta
F(\vec{k}n\downarrow)$. Then the rate equation \ref{eq:grule}
takes the form
\begin{equation}
\frac{dF(\vec{k},n,\downarrow)}{dt} =-\delta F(\vec{k}n\downarrow)
/\tau(\vec{k},n,\downarrow)
\label{eq:rate}
\end{equation}
\begin{eqnarray}
\lefteqn{1/\tau(\vec{k},n,\downarrow)= \frac{2\pi}{\hbar
N}\sum_{\vec{Q}n^{\prime}b} J^2 S p^{\prime}(\uparrow)} \nonumber
\\ &\times&\left\{\delta(\epsilon-\epsilon^{\prime}-\omega)
[n+1-f^{\prime}]
 +\delta(\epsilon-\epsilon^{\prime}+\omega)
[n+f^{\prime}] \right\}
\label{eq:tau}
\end{eqnarray}
Except for the factor $p^{\prime}(\uparrow)
=p(\vec{k}+\vec{Q}n^{\prime}\uparrow)$,
this magnon-limited scattering rate is a
perfect analog of the usual phonon-limited quasiparticle relaxation
rate from Migdal theory.
Eq. (\ref{eq:tau}) can equally well be derived by evaluation of
the Feynman diagram Fig. \ref{fig:feynman}, analytic continuation of
Matsubara frequencies $i\omega_{\nu}$ to complex frequency $z$,
and use of $1/\tau=-2{\rm Im}\Sigma(z\rightarrow \epsilon+i\delta)$.
The energies $\epsilon$ and $\omega$ used to evaluate Eq. (\ref{eq:tau})
are the experimental quasiparticle and magnon energies, which
is mandated by the occurrence of the full, not bare, Green's functions
in the self-energy graph of Fig. \ref{fig:feynman}.

\section{evaluation of relaxation rate}

Reasonable simplifying assumptions yield a ``zero parameter''
estimate of Eq. (\ref{eq:tau}) for $1/\tau$.  First, we specialize
to a state $(\vec{k},n)$ which is in an occupied surface band.
This state at low $T$ is dominantly spin up, but as seen
experimentally, it has a spin down component,
$p(\downarrow)\approx 0.13$. The Green's function $G$ and
self-energy are $2\times 2$ matrices in spin space. Eigenstates of
the 2 $\times$ 2 matrix
$G^{-1}$ are the spin-split quasiparticles, of which only the
lower (primarily majority spin) state is relevant. The energy
of this state will be denoted
$\epsilon_k$, with no band or spin index needed, and only a
two-dimensional wavevector $k$ rather than a three-dimensional
$\vec{k}$. Spin-resolved photoemission selectively depopulates a
single spin component of this state.  When the photoemitted
electron has spin down, there is a hole in the down-spin component
of the of the majority spin hole band.  This component (with
distribution function $F(k \downarrow)$) decays to equilibrium
with rate $1/\tau(k,\downarrow)$. We assume that
$p(k\downarrow)=p(\downarrow)$ is independent of $k$, and
similarly $p(k+Q\uparrow)=p^{\prime}(\uparrow)=1-p(\downarrow)$.
Then we have
\begin{equation}
\hbar/\tau(\vec{k},\downarrow)=\frac{\pi p^{\prime}(\uparrow)}
{2S}(2JS)^2 {\cal D}
\label{eq:tau1}
\end{equation}
where ${\cal D}$ is the relevant density of decay channels.  At
$T=0$, ${\cal D}$ is
\begin{eqnarray}
{\cal D}&=&\frac{1}{N} \sum_{\vec{Q}} \left\{ \right\}
=\frac{\sqrt{3}a^2c/2}{(2\pi)^3} \nonumber \\
&\times& \int d^2Q dQ_z
\left\{ \delta(\epsilon_k-\epsilon_{k+Q}-\omega(Q,Q_z))
\theta(+\epsilon_{k+Q}) \right. \nonumber \\
&& \hspace{0.5in} \left. + \delta(\epsilon_k-\epsilon_{k+Q}+\omega(Q,Q_z))
\theta(-\epsilon_{k+Q}) \right\}
\label{eq:D1}
\end{eqnarray}
We assume that the
spin waves retain their bulk character.  Therefore, although electron energies
depend only on the two-component $k$,
the spin-wave energy $\omega(Q,Q_z)$ depends
on all three components of wavevector, the $z$ direction being normal
to the surface.  The first delta function in Eq. (\ref{eq:D1}) is a
process where an empty state $k+Q$ lying above the Fermi surface is
filled from a state $k$ which scatters into it
by spin-wave emission.  In other words,
it refers to decay of an electron lying in a state $k$
which is above the Fermi energy
by at least a spin-wave energy.  The second delta function
is a process where the filled state $k+Q$ lying below the Fermi surface
scatters into an empty state $k$ by spin-wave emission; in other words,
it describes decay of a hole in the state $k$ (below the Fermi energy
by at least a spin-wave energy.)  It is only
this second process which is seen in the photoemission
data.  Let us also assume that the photohole state $k$ lies below
the Fermi energy by at least 25 meV, the maximum spin-wave energy, in
which case $\theta(-\epsilon_{k+Q})$ is guaranteed to be 1.  The
$Q_z$ integration then gives
\begin{equation}
\frac{c}{2\pi}\int_{-\pi/c}^{\pi/c}
 dQ_z \delta(\epsilon_k-\epsilon_{k+Q}+\omega(Q,Q_z))
 = \nu(k,k+Q)/s(Q)
\label{eq:zint}
\end{equation}
where $s(Q)$ is the normalized spin-wave slope
\begin{equation}
s(Q)= \frac{2\pi}{c}\left|\frac{\partial\omega}{\partial Q_z}
\right|_{\omega^{\ast}}
\label{eq:slope}
\end{equation}
and $\nu(Q)$ is the number of spin-wave states $(Q,Q_z)$ with
fixed $Q$, but any value of $Q_z$, which conserve
energy, {\sl i.e.} which satisfy
$\omega(Q,Q_z)=\omega^{\ast}=\epsilon_{k+Q}-\epsilon_{k}$.  Examining
the measured dispersion curves \cite{Koehler}, the normalized slope
can lie between 0 and a maximum which is not very different from
$\omega_{\rm max}$=25 meV.  If $\epsilon_{k+Q}-\epsilon_{k}$
is not greater than $\omega_{{\rm max}}$, then there is a fairly
good chance (something like 50\% probability) that there are 2
values, $\pm |Q_z|$, such that
the spin-wave state $(Q,Q_z)$
obeys energy conservation.  In other words, $\nu(Q)$ can be expected
to take the value 2 or 0 with about equal probability provided
$0<\epsilon_{k+Q}-\epsilon_{k}<\omega_{\rm max}$, and is definitely
0 elsewhere.

The remaining integral is
\begin{equation}
{\cal D}=\frac{\sqrt{3}a^2/2}{(2\pi)^2}\int d^2 Q \nu(k,k+Q)/s(Q).
\label{eq:D2}
\end{equation}
This integral should be evaluated using the correct dispersion
relations for the surface electron state and for the spin-wave
states, with allowance for boundary conditions and wave-function
amplitudes altering the matrix elements.  However, this requires a
large ``first-principles'' calculation of uncertain reliability.
Therefore, a slightly cavalier estimate is in order. Assume that
the surface state has parabolic dispersion $\epsilon_k=
-\epsilon_0-\hbar^2 k^2/2m^{\ast}$. Experimentally this state is
seen \cite{Fedorov} to disperse downwards in energy with effective
mass $m^{\ast} \sim 1.2m$, where $m$ is the electron mass.

\begin{figure}[t]
\centerline{\psfig{figure=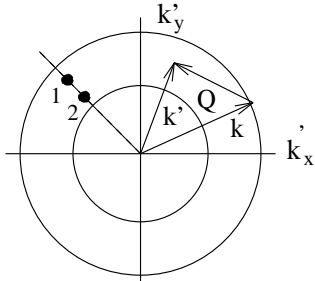,height=1.5in,width=1.61in,angle=0}}
\caption{For a given hole state $k$, the annular region shows
possible lower energy holes (higher energy filled states) which
the hole can scatter into by spin-wave emission.  The outer circle
denotes states with energy $-\epsilon_0 -\hbar^2 k^2/2m^{\ast}$,
and the inner circle denotes states higher in energy by $\omega_{\rm max}$.}
\label{fig:circle}
\end{figure}

The inequality
$0<\epsilon_{k+Q}-\epsilon_{k}<\omega_{\rm max}$ is obeyed in an
annular region shown in Fig. \ref{fig:circle}.  The area of this
region is $\pi(k^2-k^{\prime 2})$ and
$(\hbar^2 /2m^{\ast})(k^2-k^{\prime 2})=\omega_{\rm max}$.  Inside this
region we assume $\nu$ has an average value of 1 and the normalized
slope has an average value $\omega_{\rm max}/2$.  Then the density
of decay channels is
\begin{eqnarray}
{\cal
D}&=&\frac{\sqrt{3}a^2/2}{(2\pi)^2}\pi\frac{2m^{\ast}}{\hbar^2}
\omega_{\rm max} \frac{2}{\omega_{\rm max}} =
\frac{\sqrt{3}}{2\pi} \frac{m^{\ast}a^2}{\hbar^2} \nonumber \\
&\sim& 0.57 {\rm (eV)}^{-1} \label{eq:D3}
\end{eqnarray}
This is $\epsilon$-independent because of the 2-dimensionality of
the surface band.  A more careful treatment, yielding exactly the
same result, is in the Appendix.

Finally we evaluate the decay rate Eq. (\ref{eq:tau1}) by choosing
$2JS$ to be the spin splitting of the surface state, $\sim 0.65$
eV \cite{Weschke}, $S$ to be 7/2, and the fractional up-spin
probability $p^{\prime} \sim 0.87$ \cite{Fedorov}. These assumptions yield
$\hbar/\tau \sim 0.10$ eV for all photohole states $k$ which are
not too much closer than 25 meV to the top of the surface state
band.  For states closer to the top, the decay rate should
diminish because of reduction of the number of decay channels
${\cal D}$.  It is also assumed that $T$ is fairly low.  At higher
$T$, the rate is enhanced by thermal spin-wave population, but
diminished by diminishing fractional up-spin probability
$p^{\prime}(\uparrow)$. The estimate $\hbar/\tau \sim 0.10$ eV
agrees with experiment \cite{Fedorov} to greater precision 
than the uncertainties of the model.  The
calculation supports the interpretation that the source of
minority spin line-broadening is spin-flip decay.

\par
\par

\acknowledgements
I thank A. Abanov, I. Aleiner and P. Johnson for dicussions.
This work was supported by NSF grant no. DMR-0089492.

\appendix
\section*{}

\begin{figure}[t]
\centerline{\psfig{figure=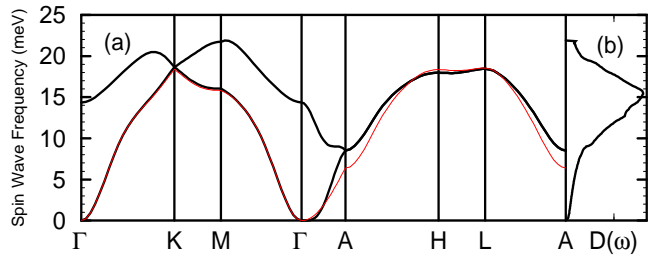,height=1.5in,width=3.5in,angle=0}}
\caption{Spin wave dispersion in Gd, calculated from
Lindgard's parameters \protect\cite{Lindgard}.  The magnon
density of states is shown in part (b) of the panel.  The thin
line represents the dispersion curve obtained by truncating
farther than first neighbors in the $z$ direction.} \label{fig:disp}
\end{figure}

Here I try to illuminate some of the approximations in the model
and also to demonstrate the surprising reliability of the estimate
\begin{equation}
\hbar/\tau=\frac{\sqrt{3}}{4}\frac{p^{\prime}(\uparrow)
m^{\ast}}{S} \left(\frac{2JSa}{\hbar}\right)^2 \label{eq:rate1}
\end{equation}
which comes from combining Eqs. (\ref{eq:tau1}) and (\ref{eq:D3}).
Spin waves in Gd have been fitted \cite{Lindgard1} with the model
\begin{equation}
\left( \begin{array}{cc} V_{11}(\vec{Q}) & V_{12}(\vec{Q}) \\
V_{12}(\vec{Q}) & V_{11}(\vec{Q}) \end{array}\right) \left(
\begin{array}{c} 1/\sqrt{2} \\ \pm 1/\sqrt{2} \end{array}\right)=
\omega_{\pm}(\vec{Q})\left(\begin{array}{c} 1/\sqrt{2} \\ \pm 1/\sqrt{2}
\end{array}\right)
\label{eq:swmatrix}
\end{equation}
where the spin wave eigenfrequencies are $\omega_{\pm}(\vec{Q})=
V_{11}(\vec{Q}) \pm V_{12}(\vec{Q})$.  The elements of the matrix
are defined by
\begin{eqnarray}
V_{11}(\vec{Q})&=& S[J_{11}(0)-J_{11}(\vec{Q})+J_{12}(0)] \nonumber\\
V_{12}(\vec{Q})&=& -SJ_{12}(\vec{Q})  
\label{eq:Vmatrix} \\
J_{11}(\vec{Q})&=& {\rm Re} \sum_{\vec{R}} J(\vec{R})e^{i\vec{Q}\cdot\vec{R}}
 \nonumber\\ 
J_{12}(\vec{Q})&=& {\rm Re} \sum_{\vec{R}} J(\vec{R}+\vec{\tau})
e^{i\vec{Q}\cdot (\vec{R}+\vec{\tau})} 
\label{eq:Jelements}
\end{eqnarray}
where $J(\vec{R})$ are fitted exchange coupling constants,
$\vec{R}$ runs over the translation vectors of the hexagonal
lattice, and $\vec{\tau}$ with $z$-component $c/2$ gives the
position of the second atom in the two-atom basis of the hcp
crystal structure.  The frequencies and density of states
calculated from this model fit experiment \cite{Koehler} well, and
are shown in Fig. \ref{fig:disp}.

Notice that the two branches $\omega_{\pm}$ are degenerate at the top of the
Brillouin zone $Q_z=\pi/c$ (directions $A\rightarrow H\rightarrow L
\rightarrow A$ in Fig \ref{fig:disp}.)
These branches can be ``unfolded'' into a zone twice as large in the
$z$-direction; $\omega_+$ is the extension of $\omega_-$
under the mapping $\omega_+(Q_x,Q_y,Q_z)=\omega_-(Q_x,Q_y,2\pi/c-Q_z)$.

\par
\par

An accurate analytic approximation for Eq. (\ref{eq:D2}) is possible
provided Lindgard's fitting parameters, Eq. (\ref{eq:Jelements})
are slightly simplified by truncating off farther than first neighbor
planes in the $z$ direction.  The alteration of $\omega(\vec{Q})$
caused by this approximation is shown in Fig. \ref{fig:disp}. 
The change is fairly small, and could be largely compensated by a
further tuning of the nearer neighbor couplings.
With this approximation, the one extended
magnon branch $\omega=\omega_{-}$ has frequency given by
\begin{equation}
\omega(\vec{Q})=a(Q_x,Q_y) + b(Q_x,Q_y) \cos(Q_z c/2).
\label{eq:freq}
\end{equation}
Then the normalized slope $s(Q)$ is given by
\begin{equation}
s(Q_x,Q_y)= \pi \sqrt{b^2-(\omega-a)^2}.
\label{eq:slope1}
\end{equation}

The remaining integral Eq. (\ref{eq:D2}) can be considered an
integral over $d^2 k^{\prime}$ running over the annulus of
Fig. \ref{fig:circle},
\begin{equation}
{\cal D}=\frac{\sqrt{3}a^2/2}{(2\pi)^2}\int_{k_1}^{k_2} d\phi
k^{\prime}dk^{\prime} \frac{2}{\pi \sqrt{b^2-(\omega-a)^2}}
\label{eq:D4}
\end{equation}
where $\omega=\epsilon(k)-\epsilon(k^{\prime})$ is the energy of
the spin wave with wavevector $(k_x-k^{\prime}_x,k_y-k^{\prime}_y,Q_z)$
which scatters the hole out of state $k$ into state $k^{\prime}$.
For a given two-vector $(Q_x,Q_y)=k-k^{\prime}$, one searches
over $Q_z$ to find whether there is an energy conserving solution.
Either there are no energy conserving spin-wave states, or else there are
$\nu=2$ such solutions at $\pm Q_z$.  Consider a path in $k^{\prime}$-space
at fixed azimuth $\phi$ shown as a dotted line in Fig. \ref{fig:circle}.
The outer circle is states $k^{\prime}$ which are degenerate in energy
with the starting state $k$.  Unless $k^{\prime}$ is the same as $k$,
there is no zero-energy spin wave which can couple these states.
Moving down the dotted line to higher energy electron states which
can fall into the hole at $k$ by magnon emission, one finds the
state labelled 1 which has just enough energy difference that an
energy conserving magnon transition is found.  The magnon has
$Q_z=0$ because it is the least energy magnon allowed to couple
on the dotted line.  Moving farther down the dotted line, one 
comes eventually to the state labelled 2 which is the highest
energy electron state which can fall into the hole in state $k$
by magnon emission.  The magnon has $Q_z=2\pi/c$ because it is
the highest energy magnon.  For both these extreme states,
$\omega=a\pm b$, the slope $s$ is zero (the magnon energy is
quadratic in $Q_z$ near $Q_z=0$ and $2\pi/c$.)  The integrand of
Eq. \ref{eq:D4} diverges at the two end points.  However, it is
an integrable divergence, and in fact, almost exactly independent
of the variables $a$ and $b$.  We observe that in going from states
1 to 2 along the dotted line, the components $Q_x$ and $Q_y$ are
not changing much, which allows us to set $a(Q_x,Q_y)$ and
$b(Q_x,Q_y)$ to constants during the $k^{\prime}$-integration.  
Then we have
\begin{eqnarray}
{\cal D}&=&\frac{\sqrt{3}a^2/2}{(2\pi)}\frac{m^{\ast}}{\hbar^2}
\int_{\epsilon-a-b}^{\epsilon-a+b} d\epsilon^{\prime}
\frac{2/\pi}{\sqrt{b^2-(\epsilon-\epsilon^{\prime}-a)^2}} \nonumber \\
&=&\frac{\sqrt{3}a^2}{(2\pi)}\frac{m^{\ast}}{\hbar^2}.
\label{eq:D6}
\end{eqnarray}
This is exactly Eq. (\ref{eq:D3}).

Finally it is appropriate to mention the hidden assumptions
about surface and bulk states.  It is implicitly assumed that
the electron surface state has a wavefunction of unit amplitude
on the surface layer and zero amplitude elsewhere.  This kind
of state will be absolutely insensitive to the $z$-component
of the magnon wave-vector, and the electron-magnon matrix
element will be $J[(2S/2N)p(\uparrow)p(\downarrow)]^{1/2}$ 
for all $k$ and $k^{\prime}$.
Suppose instead a surface electron wavefunction of amplitude $1/\sqrt{2}$ on 
each of the top two layers.  This state will couple to the
$\omega_-$ branch of magnons with the full matrix element
$J(2S/2N)^{1/2}$, but will not see the $\omega_+$ branch.
Conversely, a surface wavefunction which has amplitude
$\pm 1/\sqrt{2}$ on the top two layers, with a sign change,
will couple only to the $\omega_+$ branch and not to the
$\omega_-$ branch.  It is not reasonable that the net coupling
to magnons should depend much on the depth or details of
the surface electron wavefunction;  therefore we should ask where has the
missing magnon coupling gone when the electron
wavefunction extends two layers down instead of one.  The answer is 
in the orthogonal electron wavefunction on the
top two layers, with opposite phase relation between layer
1 and 2.  Since by assumption there is only one surface state, the orthogonal
state is not an eigenstate but a superposition of bulk states.
The missing magnon coupling is from the surface state into these
bulk states.  For the actual Gd surface state, how much of the
magnon-induced scattering is to bulk and how much is to surface
states is an unknown element.  The extreme model used here
hides this problem.  The justification is belief that
the net scattering probability should have a tendency to be conserved,
{\sl i. e.} to be weakly dependent on depth.  Similarly, we
have not asked what is the nature of the magnon states near the
surface, but instead assumed that we can use bulk magnon states.
Instead, it might be that a surface band of magnon states grabs
all the spectral weight.  Then details would be quite different,
but over-all coupling strength should be similar.



\begin{references}

\bibitem{Fedorov}       A. V. Fedorov, T. Valla, F. Liu, P. D. Johnson,
                        M. Weinert, and P. B. Allen,
                        preprint.

\bibitem{Skriver}	H. L. Skriver and I. Mertig,
			Phys. Rev. B {\bf 41}, 6553 (1990).

\bibitem{Mills}	Surface spin waves are reviewed by D. L. Mills, in
	{\sl Surface Excitations}, edited by V. M. Agranovich and
	and R. Loudon (North-Holland, Amsterdam, 1984) Ch. 3.

\bibitem{Zener}  C. Zener, Phys. Rev. {\bf 82}, 403 (1951);
	P. W. Anderson and H. Hasegawa, Phys. Rev. {\bf 100}, 675 (1955);
	P.-G. de Gennes, Phys. Rev. {\bf 118}, 141 (1960);
	K. Kubo and N. Ohata, J. Phys. Soc. Jpn. {\bf 33}, 21 (1972).

\bibitem{Millis} A. J. Millis, P. B. Littlewood, and B. I. Shraiman,
	Phys. Rev. Letters {\bf 74}, 5144 (1995);
	D. I. Golosov, Phys. Rev. Letters {\bf 84}, 3974 (2000).

\bibitem{Chattopadhyay} A. Chattopadhyay, A. J. Millis, and S. Das Sarma,
            cond-mat/0004151.

\bibitem{Lindgard}  P. A. Lindgard, B. N. Harmon, and A. J. Freeman,
            Phys. Rev. Letters {\bf 35}, 383 (1975).

\bibitem{Koehler}   W. C. Koehler, H. R. Child, R. M. Nicklow, H. G.
            Smith, R. M. Moon, and J. W. Cable,
            Phys. Rev. Letters {\bf 24}, 16 (1970).

\bibitem{Perlov}  A. Y. Perlov, S. V. Havilov, and H. Eschrig,
		Phys. Rev. B {\bf 61}, 4070 (2000).

\bibitem{Hertz}     J. A. Hertz, K. Levin, and M. T. Beal-Monod,
            Solid State Commun. {\bf 18}, 803 (1976);
            I. Grosu and M. Crisan,
            Phys. Rev. B {\bf 49}, 1269 (1994);
            M. H. Sharifzadeh Amin and P. C. E. Stamp,
            Phys. Rev. Letters {\bf 77}, 3017 (1996).

\bibitem{Reizer} V. S. Lutovinov and M. Yu. Reizer,
	Sov. Phys. JETP {\bf 50}, 355 (1979).

\bibitem{MacDonald}	A. H. MacDonald, T. Jungwirth, and M. Kasner,
	Phys. Rev. Letters {\bf 81}, 705 (1998);
	N. Furukawa, J. Phys. Soc. Jpn. {\bf 69}, 1954 (2000);
	M. J. Calder\'on and L. Brey, cond-mat/00110312.

\bibitem{Weschke}   E. Weschke, C. Sch\"ussler-Langeheine, R. Meier,
            A. V. Fedorov, K. Starke, F. H\"ubinger, and G. Kaindl,
            Phys. Rev. Lett. {\bf 77}, 3415 (1996).

\bibitem{Lindgard1}  P. A. Lindgard, in
            {\sl Spin Waves and Magnetic Excitations},
            (edited by A. S. Borovik-Romanov and S. K. Sinha),
            North-Holland, Amsterdam, 1988; vol. 1, p287.

\end{references}
\end{document}